\documentstyle[12pt,psfig]{article}
\newcommand\bm[1]{\mbox{\boldmath$#1$}}
\newcommand{\g}{\frac{G'}{G}}

\newcommand{\p}{\frac{P'}{P}}
\newcommand{\Qg}{\frac{{G'}^2}{G^2}}

\newcommand{\Qp}{\frac{{P'}^2}{P^2}}

\newcommand{\Dp}{\frac{P''}{P}}
\newcommand{\Sp}{\Sigma_{+}}
\newcommand{\Np}{N_{+}}
\newcommand{\dt}{\bm{\partial_0}}
\newcommand{\dx}{\bm{\partial_1}}
\begin{document}
\title{A self-similar inhomogeneous dust cosmology}
\author{Gernot Haager$^\natural$ and Marc Mars$^\flat$\thanks{Also
at Laboratori de F\'{\i}sica Matem\`atica,
Societat Catalana de F\'{\i}sica, IEC, Barcelona.} \\
$\natural$ Institute of Theoretical Physics,
Friedrich--Schiller--University Jena, \\
Max--Wien--Platz 1, 07743 Jena, Germany \\
$\flat$ School of Mathematical Sciences, Queen Mary
and Westfield College, \\
Mile End Road, London E1 4NS, United Kingdom.}
\maketitle
\begin{abstract}
A detailed study of an
inhomogeneous dust cosmology contained
in a $\gamma$-law family of perfect-fluid metrics recently presented
by Mars and Senovilla is performed. The metric is shown to be
the most general orthogonally transitive, Abelian, $G_2$ on $S_2$ solution
admitting an additional homothety such that the self-similar
group $H_3$ is of Bianchi type VI and the fluid flow is tangent
to its orbits. The analogous cases with Bianchi types I, II, III, V, VIII
and IX are shown to be impossible thus making this metric
privileged from a mathematical viewpoint.
The differential equations determining the metric
are partially integrated and the
line-element is given up to a first order differential
equation of Abel type of first kind and two quadratures.
The solutions are qualitatively analyzed
by investigating the corresponding autonomous dynamical system. 
The spacetime is regular everywhere
except for the big bang and the metric is complete both into the future
and in all spatial directions.
The energy-density is positive, bounded from above at any instant of time and
with an spatial profile (in the direction of inhomogeneity)
which is oscillating with a rapidly decreasing amplitude.
The generic asymptotic behaviour at spatial infinity 
is a homogeneous plane wave. 
Well-known dynamical system results indicate that this metric 
is very likely to describe the asymptotic behaviour in time
of a much more general class of inhomogeneous $G_2$ dust cosmologies. 
\end{abstract}
\newpage
\section{Introduction}
Solutions of Einstein's field equations
for dust with the energy-momentum tensor 
\begin{eqnarray}
T_{\alpha\beta} = \rho u_{\alpha}u_{\beta}, \label{Tdust}
\end{eqnarray}
where $\rho$ and $\vec{u}$ are the energy-density and
the velocity vector of the fluid, respectively,
are adequate for describing  cosmological models at late
times, i.e. when the electromagnetic radiation becomes
dynamically negligible and the evolution of the universe
is dominated by matter. The standard model describing the
geometry of the universe at late times
is the spatially homogeneous and isotropic
Friedman-Lema\^{\i}tre-Robertson-Walker
metric with vanishing pressure. However, since the universe is
not exactly isotropic and homogeneous, it becomes necessary
to consider also less symmetric dust cosmologies in order to describe those
irregularities. The first natural
generalization  are
the spatially homogeneous
Bianchi (including locally rotationally symmetric - LRS-)
cosmologies 
\cite{KSMH}
(see also \cite{WE} for more recent
developments).

The next step forward is generalizing the models so as to describe
inhomogeneous cosmologies. The simplest case here is the class
of inhomogeneous models admitting a three-dimensional
group of isometries acting on spacelike two-surfaces ($G_3$ on $S_2$
models).
That class of dust solutions includes
the spherically symmetric case (Tolman--Bondi metric)
 first found by Lema\^{\i}tre
\cite{Lema} and Tolman \cite{Tol}
(see also Krasi\'{n}ki's review book \cite{Kras97} 
for a comprehensive discussion on this topic) and
the plane and hyperbolic symmetric ones
 first discovered by Ellis \cite{Ellis67}.
The study of more general inhomogeneous models has been much
less systematic
and only very partial and incomplete results are known. Among them, we find
the remarkable Szekeres \cite{Sze75} class of inhomogeneous
irrotational dust solutions
with Petrov type D and vanishing magnetic part
of the Weyl tensor
(along the fluid congruence).
A detailed analysis of the
physical and mathematical aspects of this class of solutions 
can be found in \cite{Kras97} and references therein.

In a systematic approach for investigating inhomogeneous dust cosmologies,
the next step is the study of spacetimes
admitting two linearly independent spacelike Killing vector fields,
the so-called $G_2$ cosmologies. Very few solutions are known within this
class. The first one is contained in a $\gamma$-law family of solutions
obtained recently in \cite{Mars97}. The solution was given up to a coupled 
system of non-linear, ordinary differential equations. As we shall
see below, this metric has interesting mathematical and physical 
properties and will be carefully studied in this paper. Very recently, the
first completely explicit $G_2$ dust solutions have been obtained 
\cite{Raul97}. 
It is worth mentioning here that the situation is completely different when the
spacetime is stationary and axially symmetric (so that the isometry
group is still two-dimensional and Abelian, but acting on timelike
surfaces).  
See
\cite{Isl} for a review of dust solutions for this case.

In this paper we study in detail the inhomogeneous dust solution of Einstein's
field equations contained in the $\gamma$-law family of
perfect fluids given in \cite{Mars97}.
Its maximal isometry group is 
two-dimensional, Abelian and acting orthogonally transitively on
spacelike orbits. The metric is non-diagonal and turns out to be 
self-similar, i.e. the spacetime possesses also a homothetic Killing vector
(see  \cite{Eard} for a definition).
The self-similar group $H_3$
(spanned by the Killing vector fields and the
homothetic Killing vector) acts on
timelike hypersurfaces and the fluid velocity
is tangent to its orbits (we will use the term
{\it tangent self-similar} solution to denote those solutions for which
the fluid velocity is tangent to the self-similar orbits).
As we shall see below, this metric is the
most general orthogonally transitive $G_2$ tangent self-similar solution
such that the $H_3$ is of Bianchi type VI. Furthermore, no
analogous solutions for Bianchi types I, II, III or V exist.
Since Bianchi types VIII and IX are impossible (they do not contain an
Abelian two-dimensional
subgroup), the only remaining cases which could in principle admit
self-similar tangent solutions are Bianchi types IV or VII (these cases
 are not considered in this paper). 
As explained  below, the tangent
self-similar inhomogeneous models are known
to be suitable
candidates to describe the behaviour of more general cosmologies at
late times or near the big bang singularity.
Therefore, the solution studied here is very 
likely to describe the  asymptotic behaviour in time
of a large class of general  inhomogeneous $G_2$ dust cosmologies. 
The metric is algebraically
general, sharing this property with the explicit solutions recently
found in \cite{Raul97}
(both Szekeres and $G_3$ on $S_2$ are of Petrov type D). The system of
coupled, ordinary differential equations is partially solved so that
the metric can be given up to 
one first order ordinary differential equation of Abel
type and two quadratures.
The most relevant properties of the solution are obtained by performing
a detailed qualitative study
of the dynamical system associated with
the Abel equation. In particular, we show that the metric is free
of singularities except for the big bang, the spacetime is complete
both into the future and all spatial directions and
the energy-density is positive and bounded
from above at each instant of time. The number of essential parameters is
two.

The plan of the paper is as follows. In section 2,  the 
line-element given in \cite{Mars97} and the
differential equations to be solved are written down. The results stating the
uniqueness of the metric are presented and a summary of known results
about the relevance of self-similar cosmologies as asymptotic states for
more general models is included. 
We then rewrite the metric so that only a first order differential
equation and two quadratures remain to be solved and we give some properties
of the spacetime, like the kinematical quantities of the fluid, the
energy-density, the Petrov type and the
Killing generators of the
self-similar group, as well as the Bianchi type of the
Lie algebra they generate.
In section 3, a detailed
qualitative analysis of the two-dimensional dynamical system, including
a description of the topology of the phase space, 
the set of equilibrium points and its character and a
proof of the non-existence of periodic orbits is performed.
We then consider the asymptotic behaviour of the solution at spatial infinity
which turns
out to be a homogeneous vacuum plane wave of Petrov type N.
The coordinate transformation which brings
this plane wave metric into the standard form
is then given, thus allowing us to partially interpret 
the original coordinates in which the dust
metric is written. Finally, the limiting cases for the family
are analyzed. In section 4 we describe how this dust metric
fits
into the general framework developed by Hewitt and Wainwright
to
describe Abelian $G_2$ perfect-fluids with a $\gamma$-law equation
of state. This is particularly interesting due to the 
special character of this solution as an equilibrium point of the
dynamical system in terms of expansion normalized, dimensionless
variables.

\section{General dust solution}

Finding perfect-fluid solutions of Einstein's field equations
for $G_2$ Abelian cosmologies is a difficult task and
several simplifications have been used in order to obtain 
particular solutions. One of the most fruitful ans{\"a}tze
has been assuming that the metric
coefficients in comoving coordinates are products of functions
of only one variable (the so-called separable cosmologies).
Within the class $A(ii)$ in Wainwright's classification 
\cite{Wain81}
for $G_2$ Abelian cosmologies, the separable case with a
$\gamma$-law equation of state (i.e. $p=\gamma \rho$ where $p$ is the
pressure of the fluid) has been studied in \cite{Wils}. From  results
in that paper it is easy to infer that the particular case  $\gamma=0$
implies vacuum and so no dust solutions is possible.
The separable diagonal case was
studied in \cite{RS} (see
also \cite{Goo} for the stiff fluid solutions and the pioneering
work in \cite{WaiGoo} where
the solutions with the three-spaces orthogonal to the fluid
velocity being conformally flat were found). Again, no dust solutions
are contained in this family.
The next case in complexity (class B(i) in Wainwright's classification)
has been recently analyzed in \cite{Mars97} where 
the complete list of solutions
has been given (for some families a set of ordinary
differential equations remains to be solved). One of these
families describes a perfect fluid with $\gamma$-law
equation of state. The  particular case $\gamma=0$ (dust)
is now possible. From the
results in \cite{Mars97} it is not difficult to see that the most
general dust solution for spacetimes admitting an Abelian
maximal two-dimensional
isometry group acting orthogonally transitively on spacelike surfaces
with separable metric coefficients in comoving coordinates can be
written in the form
\begin{eqnarray}
ds^2 =  e^{2 b \hat{t}} \left( -d\hat{t}^2+d\hat{x}^2 \right)+
G P e^{(b+1) \hat{t}} dy^2 + \frac{G}{P} e^{(b-1) \hat{t}}
\left( dz+ W e^{\hat{t}} dy\right)^2, \label{Met2}
\end{eqnarray}
where $b$ is an arbitrary non-vanishing constant
and $G$, $P$ and $W$ are functions of $\hat{x}$ which must satisfy the
ordinary system of differential equations
(prime denotes derivative with respect to $\hat{x}$)
\begin{eqnarray}
b \g - \p -  \frac{W W'}{P^2} = 0,
\nonumber\\
\g \p + \Dp - \Qp + \frac{{W'}^2}{P^2} - \frac{W^2}{P^2} - b =0,
\label{seteq2} \\
\Qg - \Qp - \frac{{W'}^2}{P^2}  - \frac{W^2}{P^2} + b^2 -1=0. \nonumber
\end{eqnarray}
The symmetry assumptions in deriving this metric
involved only a maximal isometry group of dimension two. However, it turns
out that the metric (\ref{Met2})
possesses also a homothetic Killing vector (in fact, this also holds
for the full class of solutions with $\gamma$-law equation of state in
\cite{Mars97}).
The three-dimensional homothety group $H_3$ belongs to the Bianchi type VI
and its orbits are timelike everywhere with the the fluid flow tangent to them.

Spacetimes admitting a three-dimensional homothety group have been
investigated by several authors in the last few years
(see e.g. \cite{Carot96} and references therein).
In particular, a recent paper
by Carot {\it et al} \cite{Carot94}
gives the explicit time dependence for this kind of metrics.
Furthermore, in \cite{Carot96} it is stated that
tangent self-similar perfect fluids of Bianchi types I and II
have to satisfy $\rho + 3 p = 0$ thus excluding dust solutions.
Combining their results
with the uniqueness of the metric (\ref{Met2}) as the only dust
solution with separable
coefficients in comoving coordinates,
the following proposition holds

\vspace{5mm}
{\it The line-element (\ref{Met2}) satisfying the equations (\ref{seteq2})
is the most general dust solution of Einstein's field equation for
spacetimes of class B in Wainwright's classification of inhomogeneous
cosmologies, admitting one homothety and such that the fluid flow is tangent
to the self-similar group orbits and
the Bianchi type of the three-dimensional homothetic group is VI. Furthermore,
the analogous cases in which the self-similar group is
of Bianchi types I, II, III and V are impossible.}

\vspace{5mm}
Additionally, $G_2$ Abelian cosmologies with an additional
homothetic vector field such that the three-dimensional self-similar group
$H_3$ is of Bianchi types VIII and IX are clearly impossible because these
Bianchi types do not contain Abelian two-dimensional subalgebras.
In principle, $G_2$ dust solutions with an $H_3$ of 
Bianchi types IV and VII could admit tangent solutions
but none of these algebras
admit diagonal metrics (unless the spacetimes possess more than two
isometries). Furthermore, the diagonal limit of the metric (\ref{Met2})
(obtained for $W\equiv 0$) can be trivially seen
to be either Minkowski or a Bianchi cosmology.
Thus, the following result also holds

\vspace{5mm}
{\it No dust solutions of Einstein's field equations exist for
diagonal inhomogeneous cosmologies (B(ii) in Wainwright's classification)
admitting a homothetic symmetry such that the fluid flow is
tangent to the self-similar orbits.}
\vspace{5mm}

Tangent self-similar cosmologies
play a prominent role in describing
the asymptotic behaviour (either near the big bang singularity
or at $t \rightarrow \infty$) for more general cosmological
models. This result is one of the main consequences of the
dynamical system analysis of cosmological models which has been
undertaken in recent years (see
\cite{WE} and references therein).
The dynamical system approach
consists of writing Einstein's field equations as
first order partial differential equations for some
kinematical quantities associated with an orthonormal
tetrad which must be chosen appropriately
\cite{Ellis71}.
Then, an expansion normalized set of variables is introduced
and a qualitative analysis of the resulting dynamical system
is performed. This approach for analyzing
cosmological models started by considering the 
simplest cases, namely homogeneous cosmologies in which
the Einstein field equations become ordinary differential
equations and therefore the associated phase space
is finite dimensional. The mathematical
theory for describing finite dimensional dynamical
systems is 
well-developed and a good number of properties of the solutions
can be derived without the need of solving the
equations explicitly. 
One of the most relevant results which
were proven in that context \cite{Wain89}, \cite{Wain93}
is that all equilibrium points of the dynamical system
associated with non-tilted Bianchi cosmologies (i.e. such that
the fluid velocity is orthogonal to the group orbits)
satisfying a $\gamma$-law equation of state 
correspond to
self-similar cosmologies, i.e. spatially homogeneous
spacetimes admitting a homothetic Killing vector.
Thus, this class of self-similar
Bianchi models plays a relevant role as asymptotic states
for more general cosmologies.

For the case of $G_2$ cosmologies, the phase space becomes, in general,
infinite dimensional since the tetrad form of Einstein's equations are 
first order {\it partial} differential equations. The mathematical
theory in this case is less developed and the 
results which can be obtained are more restricted.
However, Hewitt and Wainwright \cite{HeW} reformulated the field equations
as a dynamical system for orthogonally transitive $G_2$ 
perfect fluid cosmologies with a $\gamma$-law equation of state.
In \cite{HeW}
it has  been proven 
that the
equilibrium points of that dynamical system
 also correspond to self-similar models.
They admit a three-dimensional self-similar group $H_3$ (spanned
by the two Killing vectors and one homothetic Killing
vector) acting on timelike hypersurfaces. The fluid velocity
vector is tangent to the three-dimensional similarity
orbits. These two results suggest that tangent self-similar models
are important in describing the asymptotic behaviour of
a large class of inhomogeneous cosmologies. Thus,
the unique (for the Bianchi types I, II, III, V and VI of the
$H_3$ group) tangent self-similar solution (\ref{Met2})
is very likely to be relevant for describing the asymptotic state
of general dust cosmologies.

All these results above show that the metric (\ref{Met2})
is  privileged from a mathematical point of view, and, thus,
it deserves a more detailed analysis. Furthermore,
as we shall see below,
it turns out that the metric is also well-behaved
from a physical viewpoint.
Let us then proceed with the analysis of the metric
(\ref{Met2})
and field equations (\ref{seteq2}). First of all, it is
immediate to see that
the transformation
\begin{eqnarray*}
b \longleftrightarrow -b, \quad \hat{t} \longleftrightarrow - \hat{t},
\quad
z \longleftrightarrow y, \quad P \longleftrightarrow \frac{P}{P^2+W^2},
\quad
W \longleftrightarrow \frac{W}{P^2+W^2}
\end{eqnarray*}
leaves the metric (\ref{Met2}) invariant.
Thus, the constant $b$ can be
restricted to be positive without
loss of generality. Furthermore, the energy-density of the fluid is
well-behaved only for $b \geq 1$ (otherwise it is negative everywhere) so that
we can restrict the analysis to $b \geq 1$ and introduce
another constant $\alpha$ defined as
\begin{eqnarray*}
\sin \alpha = \frac{1}{b}, \hspace{1cm} \cos{\alpha} \geq 0.
\end{eqnarray*}
The system of differential equations we have to deal with
constitutes (after substituting $\g$ obtained from the
first equation in (\ref{seteq2}) into the other two equations)
a non-linear coupled system of
two ordinary differential equations for two unknowns. One of the
equations is of second order and linear
in the highest derivative and the other is of the first order
but quadratic
in the highest derivatives. Thus, the system is very complicated
and it must be simplified. To that end we rewrite the metric, after
redefining the coordinates $\hat{t}$ and $\hat{x}$ and rescaling $y$
and $z$, as
\begin{eqnarray*}
ds^2 = -dt^2 + t^2 \frac{dx^2}{L(x)} +GP t^{1+\sin\alpha} dy^2 +
\frac{G}{P} t^{
1- \sin \alpha} \left (dz + W t^{\sin{\alpha}} dy \right )^2
\end{eqnarray*}
where $L(x)$ is an arbitrary non-vanishing function at
our disposal.
This function has to be chosen appropriately in order  to simplify
the field equations without introducing unnecessary coordinate singularities
in the metric. This is not a trivial task and a careful investigation
of the equations (\ref{seteq2}) is involved. It turn out that the
choice
\begin{eqnarray}
L(x) = \sin^2{\alpha} + \cos^2 \alpha \cos^2 \left ( \frac{H(x)}{2}  \right),
\label{ele}
\end{eqnarray}
allows for an explicit integration of the function $W$ and
decomposes the remaining field equations into a first order ordinary
differential equation for the function $H$ and two quadratures for
$G$ and $P$ respectively, which is a substantial simplification
of the problem. The metric can be written as
\begin{eqnarray}
ds^2  & = & - dt^2 + \frac{ t^2 dx^2}{\sin^2{\alpha} + \cos^2{\alpha}
\cos^2 \left (\frac{H}{2} \right)} + t^{1+\sin \alpha}
 G P dy^2  \nonumber \\
& &  \hspace{3cm}+ t^{1-\sin \alpha } \frac{G}{P} \left( dz+
 \frac{\cos \alpha}{\sin \alpha} t^{\sin \alpha}
  P  \cos \left (\frac{H}{2}
\right) dy \right)^2.
\label{metric1}
\end{eqnarray}
The first order differential equation for the function $H$ is best
analyzed when written
as an autonomous dynamical system in two dimensions.
It reads, explicitly,
\begin{eqnarray}
\frac{dH}{dx} & = & 2 \cos Q, \nonumber \\
\frac{dQ}{dx} & = & \cos Q + \frac{\sin Q \sin H}{
2 \tan^2 \alpha + 1 + \cos H }.
\label{dynode}
\end{eqnarray}
where we have introduced a new function $Q(x)$.
The change $Q \equiv \arctan \left [ M(H) \right ]$
transforms this dynamical system back into
the first order ordinary differential equation
\begin{eqnarray*}
\frac{dM}{d H} = \frac{1}{2} \left (1+ M^2 \right) \left [
1 + \frac{M \sin H}{2 \tan^2 \alpha + 1 + \cos H} \right ],
\end{eqnarray*}
which is an Abel equation of the first kind
\cite{Kamke60}. It does not belong to the known integrable cases 
\cite{Heike87}. Although apparently simple, solving
this Abel equation is still very difficult
and no particular solutions have been found. However, a qualitative analysis
of the dynamical
system (\ref{dynode}) is enough for a good understanding
of the solutions. This will be performed in the next section.

The other two metric coefficients can then be calculated by
the quadratures
\begin{eqnarray}
\frac{G_x}{G}  & = & \sin Q, \label{quadra1} \\
\frac{P_x}{P} & = & \frac{ 2 \sin \alpha \sin Q +  \cos^2 \alpha \cos Q
\sin H}{  2 \sin^2 \alpha + \cos^2 \alpha (1+ \cos H) }.
\label{quadra2}
\end{eqnarray}
The number of essential parameters in the family (\ref{metric1}) is two;
the explicit constant $\alpha$ in the line-element and the
initial condition in the dynamical system (\ref{dynode}). The
integration constants in $G$ and $P$ are superfluous and can be set
equal to one by trivial redefinitions of $y$ and $z$.

Let us now describe some of the properties of the dust
solution we are considering.
The fluid velocity vector is
$\vec{u} = \frac{\partial}{\partial t}$ and the energy-density 
reads
\begin{eqnarray}
\rho = \frac{ \cos^2 \alpha \sin^2 \left ( \frac{H}{2} \right)}{t^2}
\label{density}
\end{eqnarray}
which is positive everywhere
and bounded from above at any instant of time $t$. The
metric (\ref{metric1}) has a big bang singularity
at $t=0$, where the energy-density blows up.
Regarding the kinematical quantities, the fluid velocity is
obviously geodesic and irrotational and
the expansion is spatially homogeneous and reads
\begin{eqnarray*}
\theta = \frac{2}{t},
\end{eqnarray*}
showing that the fluid never recollapses.
The shear tensor is highly anisotropic and the shear scalar is
\begin{eqnarray*}
\sigma_{\alpha \beta} \sigma^{\alpha \beta} =
\frac{4 -3 \cos^2 \alpha \sin^2 \left ( \frac{H}{2} \right)}{6 t^2}.
\end{eqnarray*}
All fluid
elements are expanding and a comoving observer would measure
red shift in every direction. 
Regarding the Petrov type, the metric is algebraically
general (Petrov type I) except for the particular case $\alpha=
\frac{\pi}{2}$ when it degenerates to conformally flat.
In this case
the energy-density (\ref{density}) also vanishes
and the metric becomes flat.
The magnetic part of the Weyl tensor along the fluid velocity
vector in the spacetime (\ref{metric1}) is non-vanishing.
Except for some particular subcases described below,
the metric (\ref{metric1}) has the only two Killing vectors
$\vec{\xi}_2 = \partial_y$ and  $\vec{\xi}_3= \partial_z$.
The homothetic Killing vector is
\begin{equation}
\vec{\xi}_1 = t \frac{\partial}{\partial t} + \frac{\left (1- \sin \alpha \right)}{2}
 y \frac{\partial}{\partial y} + \frac{ \left (1 + \sin \alpha \right)}{2}
z \frac{\partial}{\partial z}
\label{homovec}
\end{equation}
and the three-dimensional homothetic Lie algebra spanned by these three vectors
is of Bianchi type $\mbox{VI}_h$,  where $h$ is given by
\begin{eqnarray*}
h = - \frac{1}{\sin^2 \alpha} = -b^2.
\end{eqnarray*}

This dust solution can be matched to a vacuum solution
with two commuting
spacelike Killing vectors (which can therefore be interpreted
as a gravitational wave), but the vacuum metric cannot possess
an additional homothetic Killing vector any longer. Indeed, it is easy
to see that the matching between the dust solution and any vacuum metric
must be performed across one of the homothetic group orbits in
the dust solution.
If we assume that the homothetic Killing vector extends to the
vacuum region, the time dependence of the vacuum metric becomes
determined everywhere and must coincide (due to the continuity
of the first fundamental forms on the matching hypersurface)
with the time dependence in the dust region. Thus, the vacuum metric
must be a homogeneous plane wave. However, the remaining set of matching
conditions forbids the matching of a homogeneous plane wave and the
self-similar dust cosmology at finite distances.
The field equations for the vacuum solution matching the metric (\ref{metric1})
are still quite complicated, and no explicit solution has been found.

\section{Phase space analysis and limiting solutions}
In this section we will analyze the spatial behaviour of the dust
metrics (\ref{metric1}). To this aim, we have to study the
dynamical system (\ref{dynode}). This is a two-dimensional
dynamical system, depending on an arbitrary parametric constant
$\alpha$, which
possesses the two discrete symmetries $Q \rightarrow Q + 2 \pi$ and
$H \rightarrow H + 2 \pi$. There is another
discrete symmetry $\{Q \rightarrow - Q, H \rightarrow - H\}$.
Thus, it is sufficient to examine the domain
${\cal U} = (- \pi , \pi] \times (-\pi , \pi]$ in
the phase space diagram for $H$ and $Q$
and use the above periodicity. More strictly,
the phase space of the dynamical system (\ref{dynode})
is topologically a two-dimensional torus $S^1 \times
S^1$ which can be described by the coordinates 
$\left (H,Q\right) \in {\cal U}$ by identifying the two vertical
and the two horizontal boundaries respectively.
The qualitative behaviour of the
dynamical system is independent of the value of the
parameter $\alpha$ and no bifurcations occur. A plot
for the phase space portrait with a typical value  $\alpha= \pi/4$
(which will be assumed for all Figures)
is given in Fig.1. In order to understand the behaviour of the solutions
we must consider the set of fixed points of the dynamical system.
There are four fixed points; two of them are saddle
points, one is an attractor and the remaining one is a repellor. They
are given by
\begin{eqnarray}
\mbox{Saddle points:} \hspace{5mm}
\left (Q = \frac{\pi}{2} \, , \, H = \pi \right), \quad
\left( Q= - \frac{\pi}{2} \, , \, H = \pi \right) \label{fpoint2}\\
\mbox{Attractor:} \hspace{3mm} \left (Q = \frac{\pi}{2} \, , \, H = 0 \right),
\hspace{1cm} \mbox{Repellor:} \hspace{3mm} 
\left (Q = -\frac{\pi}{2} \, , \, H = 0 \right)
\label{fpoint1}.
\end{eqnarray}
The two solutions corresponding to the two-saddle points are equivalent
and represent a
spatially homogeneous self-similar cosmology (with the three-dimensional
isometry group of Bianchi type VI). The two solutions corresponding to
the attractor and repellor are also equivalent and represent
vacuum spacetimes.
Since the homotopy group of the torus is non-trivial (it is isomorphic
to the additive group Z), any solution starting
at the repellor and finishing at the attractor can be classified by the
number of times the trajectory goes around the torus before reaching its
endpoint (i.e. they can be classified by the homotopy class they belong to).
In order to understand the possible behaviour of the solutions, it is
convenient to draw the set of trajectories finishing or starting at the
saddle points. The plot is given in Fig.2.
These curves divide the phase space diagram into basins of attraction
which give a clear description of the behaviour of the solutions. 
In particular, we can conclude from this diagram that no periodic solutions
of (\ref{dynode}) exist. This follows from the
fundamental fact that two different orbits in any
dynamical system cannot intersect anywhere and that any closed orbit which is
contractible to a point must contain an equilibrium point in its interior.
Thus, all the solutions of (\ref{dynode}), except
those starting and finishing on the saddle points, approach asymptotically
the repellor (for $x \rightarrow - \infty$) and the attractor (for
$x \rightarrow + \infty$). Regarding the different
homotopy classes the
solutions can belong to, it follows from Fig 2. that two different cases
are possible. Either the solutions
go from the repellor to the attractor without going around the torus, or
they go around it at most once. In the second case, the solutions
can belong to two different basins of attraction, which can be distinguished
by the number of times the boundary of ${\cal U}$ is crossed. This boundary
can be crossed either once (in the $H$ direction) or twice (one
in the $H$ and one in the $Q$ direction).
A typical solution for each one of 
the three possible cases is shown in Fig. 3, together with the
energy-density $\rho$, which oscillates with rapidly decreasing amplitude.
 
A numerical analysis of the solutions of (\ref{dynode})
shows that the asymptotic approach to the stable fixed points
(\ref{fpoint1}) is quite fast and that the solutions
oscillate around those points in their asymptotic approach.
Since every crossing of the
value $H=0$ corresponds to a vanishing value for the energy-density
(see (\ref{density})), it turns out that the energy-density profile of the 
solutions is oscillating in the positive range, reaching the zero 
value at a discrete
number of points and with a rapidly decreasing amplitude     
(see the energy-density plots in Fig.3 for examples of this
behaviour)

Let us now study in detail the line-elements corresponding to the
fixed points, since this will allow us for
a partial interpretation of the
coordinate system $\{ t,x,y,z \}$ in (\ref{metric1}).

The asymptotic vacuum solution is represented in the coordinate system
(\ref{metric1}) by the line-element
\begin{eqnarray}
ds^2= - dt^2 + t^2 dx^2 +\left (te^{\epsilon x} \right )^{1+\sin \alpha}
 dy^2+\left (te^{\epsilon x} \right )^{1-\sin \alpha}
 \left (dz+  \frac{\cos \alpha}{\sin \alpha}
\left (t e^{\epsilon x} \right) ^{\sin \alpha} dy
\right)^2 \label{vacinfty}
\end{eqnarray}
where $\epsilon = +1$ ($\epsilon = -1$) for $x \rightarrow + \infty$
($x \rightarrow - \infty$). This vacuum spacetime is a homogeneous
plane wave with a six-dimensional isometry group acting transitively
on $V_4$. The Petrov type is N and the only repeated  principal
null direction is $\vec{l} = t \partial_{t} - \epsilon
\partial_x$. Thus,
the inhomogeneous dust metric 
contains a matter and a radiative part.  
The radiative part dominates at large distances in the spatial
direction perpendicular to the two-planes spanned by the isometries.
In a cosmological context it could be interpreted as a gravitational
wave background superposed to the expanding dust.
This asymptotic behaviour at spacelike infinity seems to be quite common in
this kind of situations for perfect fluids.
Hewitt {\it et al} \cite{Wain88} performed
a dynamical system analysis of tangent self-similar
{\it diagonal} $G_2$ perfect fluids with a
$\gamma$-law equation of state. They concluded that the spacelike asymptotic
behaviour can be either matter dominated (i.e. with a non-zero value for the
energy-density at spacelike infinity) or vacuum dominated (in which
the energy-density tends to zero at infinity). In this second case,
the vacuum asymptotic behaviour is also a homogeneous plane wave.

The very neat behaviour of the solution
at spatial infinity can provide us with a partial
interpretation for the coordinate system $\{t,x,y,z \}$. In order to
do so, we consider the coordinate change which
brings (\ref{vacinfty}) into the standard form for plane waves
\cite{KSMH}
\begin{eqnarray}
ds^2 = -2 du dv + 2 \left [ A(u) \left (Z^2 - Y^2 \right) + 2 B(u) Y Z
\right ] du^2 +dY^2 + dZ^2. \label{standpw}
\end{eqnarray}
It can be easily seen that this coordinate transformation is
given by
\begin{eqnarray}
y = - u^{-\frac{1+\sin \alpha}{2}}
\left [Y\sin(\beta-\alpha)+ Z\cos(\beta-\alpha) \right ], \quad
z= \frac{u^{\frac{-1+\sin \alpha}{2}}}{\sin \alpha}
\left (Y\sin\beta+Z\cos\beta \right), \label{transpw} \\
t e^{\epsilon x} = u, \quad
t e^{-\epsilon x} =2 v-\frac{1}{2u} \left [ \left (Z^2-Y^2 \right )
 \sin(2\beta-\alpha)-2Y Z \cos( 2\beta-\alpha)+Y^2+Z^2 \right ],
\nonumber
\end{eqnarray}
where
\begin{eqnarray}
\beta \equiv \frac{\cos \alpha}{2} \log {u}. \label{gama}
\end{eqnarray}
The two function $A(u)$ and $B(u)$ in (\ref{standpw}) read
\begin{eqnarray*}
A(u)=\frac{\cos \alpha}{4u^2} \cos(2\beta-\alpha), \hspace{1cm}
B(u) =\frac{\cos \alpha}{4u^2} \sin(2\beta-\alpha)
\end{eqnarray*}
So, near spatial infinity a partial interpretation of the coordinate system
(\ref{metric1}) is possible. The coordinate $x$ measures a spacelike
distance along the direction of propagation of the plane wave. 
The coordinates $y$ and $z$ are
straight lines lying in a plane tangent to the null hypersurface
corresponding to the plane wave front. They are not
orthogonal lines, the angle between them being directly related to 
the parameter $\alpha$.
The lines $y= \mbox{const.}$ and $z=\mbox{const.}$ are scaled by different
units, so there is no rotational symmetry in the corresponding
plane. Moreover, the straight lines associated with $y$ and $z$
are rotated by an angle $\beta$ (which depends on $u$ and therefore
is different for different wave fronts)
with respect to the orthogonal straight lines $Y$, $Z$ 
associated with the plane of symmetry of the wave. It follows then that
the coordinates $y,z$ are rotating with respect to the homogeneous plane
wave background. 

The solution corresponding to the
saddle points is a diagonal spatially homogeneous spacetime of
Bianchi type $\mbox{VI}_{\left (-\sin^2 \alpha\right)}$. 
The line-element
(\ref{metric1}) takes the form
\begin{equation}
ds^2 = -dt^2 + t^2 dx^2 + ( t e^{x})^{1+\sin \alpha} dy^2+
       (t e^{-x})^{1-\sin \alpha} dz^2
  \label{bianchimetric}
\end{equation}
which possesses the additional killing vector field
\[ \vec{\xi}_1 = \frac{\partial}{\partial x} -\frac{1 + \sin \alpha}{2}y \frac{\partial}{\partial y} +
\frac{1 - \sin \alpha}{2}  z \partial_z
\]
and the energy density is $\rho = \frac{\cos^2 \alpha}{t^2}$. This metric
was first obtained by Collins \cite{Coll71} and appears listed in the
catalogue of self-similar solutions by Hsu and Wainwright \cite{HWain86}.
This solution also appears as a saddle point in a dynamical system
investigation of tangent self-similar, $G_2$ diagonal perfect fluids with
a $\gamma$-law equation of state \cite{Wain88}. 

For $\alpha =
\frac{\pi}{2}$ the line-elements (\ref{standpw}) and
(\ref{bianchimetric}) are exactly the same. Moreover, the
expressions for $A(u)$ and $B(u)$ vanish when 
$\alpha = \frac{\pi}{2}$ and therefore  
the metric becomes flat.
The transformation (\ref{transpw})
particularized to  $\alpha = \frac{\pi}{2}$ simply reads
\begin{eqnarray*}
y = \frac{Y}{u}, \hspace{1cm}
z=  Z,
\hspace{1cm}
t e^{x} = u, \hspace{1cm}
t e^{-x} =2 v - \frac{Y^2}{u} ,
\end{eqnarray*}
which can be rewritten using the standard orthogonal coordinates
${T,X,Y,Z}$ for Min\-kow\-ski spacetime as
\begin{eqnarray*}
t =  \sqrt{T^2-X^2-Y^2}, \hspace{1cm}
e^x =  \frac{ \left (T+X \right)}{ \sqrt{2} \sqrt{T^2-X^2-Y^2}}, \hspace{1cm}
y =  \frac{\sqrt{2}\, \, Y}{T+X}, \hspace{1cm} z  =  Z.
\label{flattransf}
\end{eqnarray*}
The four velocity ${\bm u} = -{\bf dt}$ is transformed to
\begin{eqnarray*}
\bm{u} =  -\frac{1}{\sqrt{T^2-X^2-Y^2}} \left(T {\bf dT}+ Y {\bf dY} +
X {\bf dX} \right) .
\end{eqnarray*}
Let us finish the discussion on the solution (\ref{metric1}) by considering
another limiting case. The family (\ref{metric1})
is defined for every value of $\alpha$ except
when $\sin \alpha =0$ (this corresponds to $b = \infty$ in the
metric (\ref{Met2})). It is therefore convenient to consider the limiting
case $\sin \alpha \rightarrow 0$. It can be easily seen that rescaling
the coordinates in order to reabsorb the diverging behaviour in the
coefficient $dy dz$ in (\ref{metric1}) leads necessarily
to a degenerate metric. Therefore,
the only possibility is that $\cos \left (\frac{H}{2} \right)$ tends
to zero when $\sin \alpha \rightarrow 0$. Since the metric coefficient in
$dx^2$ then diverges, it is necessary to perform a coordinate
transformation to obtain a regular metric. Thus, we define
\begin{eqnarray*}
\cos \left (\frac{H}{2} \right ) = \delta(r) \sin \alpha,
\hspace{1cm}
\frac{dx}{\sin \alpha \sqrt{1+ \delta^2}} = dr
\end{eqnarray*}
which brings both the metric and the field equations into a form
allowing the limit $\sin \alpha \rightarrow 0$. The field
equations turn out to be completely integrable and the resulting
line-element can be written as
\begin{eqnarray}
ds^2 = -dt^2 + t^2 dr^2 + \frac{t}{\cosh r} dy^2 + t \cosh r \left (
\frac{}{}
dz + \tanh r \, \,dy \right)^2, 
\end{eqnarray}
which can be transformed into the particular case $\sin \alpha=0$ in
(\ref{bianchimetric}) by the coordinate transformation
\begin{eqnarray*}
r \longrightarrow x, \hspace{1cm} \frac{y+z}{\sqrt{2}} \longrightarrow y,
\hspace{1cm} \frac{y-z}{\sqrt{2}} \longrightarrow z.
\end{eqnarray*}
Thus the solution (\ref{bianchimetric}) is the correct spatially
homogeneous limiting case for all values of $\alpha$.

As we have emphasized above, the dust model we study in this paper
corresponds to an equilibrium point in the dynamical system
approach for $G_2$ Abelian perfect-fluids developed by
Hewitt and Wainwright \cite{HeW}. It is, therefore, interesting to 
analyze how this solution fits into that framework. This
will be the task of the next section.

\section{Description of the solution using the tetrad approach}

Throughout this section, we shall adopt the same notation and definitions as in
\cite{HeW}. One of the main results in that paper
is that every  equilibrium point of the dynamical system they construct
(in terms of expansion normalized, dimensionless connection coefficients)
corresponds to an orthogonally transitive
$G_2$ perfect-fluid (assuming a linear equation of state)
which admits a homothetic vector field lying in
the three-plane spanned by $\vec{u}$ and the two Killing vectors (i.e.
a tangent self-similar solution). Since these conditions are met by the
metric we analyze, we can assume $\bm{\partial_0} X=0$ where $X$ represents
any dimensionless variable. Furthermore, the fluid is dust and hence
the four velocity
is geodesic. Imposing these conditions on the dynamical
system in \cite{HeW},
we readily obtain
\begin{eqnarray*}
q= \frac{1}{2} \hspace{1cm} \Sp= -\frac{1}{4}, \hspace{1cm} r=0,
\end{eqnarray*}
which, in particular, imply that $\dt$ and $\dx$ commute. Thus,
every tangent self-similar dust solution has constant and positive
deceleration parameter $q$ and the metric is never inflating.
The defining equations for $q$ and $r$ provide now
\begin{eqnarray*}
\Omega = \frac{3}{4} - 12 \left ({\tilde{\Sigma}_{22}}^2 + 
{\tilde{\Sigma}_{23}}^2 \right),
\hspace{1cm}
A = 12 \left (\tilde{\Sigma}_{22} \tilde{N}_{23} - \tilde{\Sigma}_{23}
{\tilde{N}_{22}} \right),
\end{eqnarray*}
indicating a very simple relationship between the 
density parameter $\Omega$ and the dimensionless shear scalar. Using all this
information, the dynamical system splits into a set of
ordinary differential equations
\begin{eqnarray}
\dx R  = -18 ( \tilde{\Sigma}_{22} {\tilde{N}_{22}} + \tilde{\Sigma}_{23}
\tilde{N}_{23}), \nonumber \hspace{25mm} \\
\dx \tilde{\Sigma}_{23} =
 -2 \left (\Np \tilde{\Sigma}_{22} + R \tilde{N}_{23} \right), 
\hspace{1cm}
\dx \tilde{\Sigma}_{22} = 2 \left (\Np \tilde{\Sigma}_{23}
- R {\tilde{N}_{22}} \right ), \nonumber\\
\dx \tilde{N}_{23} =
24 \tilde{N}_{23} ( \tilde{\Sigma}_{22} \tilde{N}_{23} - \tilde{\Sigma}_{23}
{\tilde{N}_{22}}) - 2\Np {\tilde{N}_{22}} + 2 R \tilde{\Sigma}_{23} -
\frac{3}{2}
\tilde{\Sigma}_{22}, \label{dina}\\
\dx \tilde{N}_{22} = 24 {\tilde{N}_{22}} ( \tilde{\Sigma}_{22}
\tilde{N}_{23} - \tilde{\Sigma}_{23} {\tilde{N}_{22}}) + 2\Np \tilde{N}_{23}
+ 2 R \tilde{\Sigma}_{22} + \frac{3}{2}
\tilde{\Sigma}_{23}, \nonumber
\end{eqnarray}
and the algebraic constraint
\begin{eqnarray*}
1 - 16 ({\tilde{N}_{22}}^2 + \tilde{N}_{23}^2 +\tilde{\Sigma}_{22}^2 +
\tilde{\Sigma}_{23}^2 ) + 16^2 (\tilde{\Sigma}_{23} {\tilde{N}_{22}} - \tilde{\Sigma}_{22} 
\tilde{N}_{23})^2 = 0,
\end{eqnarray*}
which is a first integral of (\ref{dina}).
It is a matter of simple calculation to check that
\begin{eqnarray}
R^2 = 9 ({\tilde{\Sigma}_{22}}^2 + {\tilde{\Sigma}_{23}}^2) + \kappa
\label{Fint}
\end{eqnarray}
(where $\kappa$ is an arbitrary constant) is also 
a first integral of (\ref{dina}).
The sign of this constant can be related to the Bianchi type of the
homothetic algebra.
It turns out that $\kappa < 0 $ corresponds
to Bianchi VI, $\kappa=0$ to Bianchi IV and $\kappa>0$ to Bianchi VII.
Thus, the dynamical system is six-dimensional (in the variables
$R$, $\tilde{\Sigma}_{22}$, $\tilde{\Sigma}_{23}$, $\tilde{N}_{22}$,
$\tilde{N}_{23}$ and $N_{+}$) with 
two polynomic constraints. In order to simplify it further
we take advantage of the rotational
freedom in the $\bm{e_2}$, $\bm{e_3}$ plane (the rotation angle $\phi$
must be
constant both along the group orbits and along the integral lines of the fluid
velocity in order to comply with all the conditions we already imposed).
Under such a transformation, R behaves
as an scalar, $N_{+}$ transforms as $N_{+} \longrightarrow  N_{+} + \dx \phi$,
$\tilde{\Sigma}_{AB}$  and $\tilde{N}_{AB}$
as  symmetric rank two tensors.
The Lie algebra we study in this paper is of
Bianchi type VI and
therefore $\kappa$ is strictly negative, which will be assumed from
now on. Hence, the first integral (\ref{Fint}) implies $R^2 < 9(
{\tilde{\Sigma}_{22}}^2 +  {\tilde{\Sigma}_{23}}^2)$, which allows
us to fix $\phi$ so that the condition
\begin{eqnarray}
R = - 3 \tilde{\Sigma}_{23} \label{choice}
\end{eqnarray}
is fulfilled.
Notice that this choice cannot be imposed in the
case of Bianchi type VII (for which $\kappa >0$). The
choice (\ref{choice}) immediately implies
$\tilde{\Sigma}_{22}= \frac{\sqrt{-\kappa}}{3}$. The non-negativity
of the energy-density demands $\frac{4 \sqrt{-\kappa}}{3} \leq
 1$ and we can set
$\frac{4 \sqrt{-\kappa}}{3} \equiv \sin \alpha$.
Furthermore, the differential equation for
$\tilde{\Sigma}_{22}$ above fixes $N_{+} = -3  \tilde{N}_{22}$.
Inserting all this information
into the dynamical system, we obtain the following set of equations
\begin{eqnarray}
\dx \tilde{\Sigma}_{23} = \frac{3}{2} \left (4 \tilde{\Sigma}_{23} 
\tilde{N}_{23} + \sin \alpha \tilde{N}_{22} \right), 
\hspace{2cm} \nonumber
\\
\dx \tilde{N}_{23} = 6 \left (\tilde{N}_{23}^2 \sin \alpha - 4 
\tilde{N}_{22} \tilde{N}_{23} \tilde{\Sigma}_{23} + \tilde{N}_{22}^2
- {\tilde{\Sigma}_{23}}^2 \right ) -\frac{3 \sin \alpha}{8}, 
\label{dina2} \\
\dx \tilde{N}_{22} = 6 \tilde{N}_{22} \left [
\left (\sin \alpha - 1 \right) \tilde{N}_{23}  - 4 
\tilde{N}_{22} \tilde{\Sigma}_{23} \right ] +
 \frac{3 \left(1-\sin \alpha \right ) 
\tilde{\Sigma}_{23}}{2}, \nonumber 
\end{eqnarray}
and the constraint 
\begin{eqnarray}
\cos^2 \alpha  \left ( 1- 16 \tilde{N}_{23}^2 \right) - 
16 \tilde{\Sigma}_{23}^2 \left ( 1- 16 \tilde{N}_{22}^2 \right)
-16 \tilde{N}_{22} \left (\tilde{N}_{22} + 8 \tilde{N}_{23} 
\tilde{\Sigma}_{23} \sin \alpha \right)=0. \label{surf}
\end{eqnarray}
Hence, the dynamical system is two-dimensional as the motion
happens on the surface $S$ defined by (\ref{surf}).
In order to obtain an adequate
description of the problem we should find appropriate 
global coordinates
on the surface $S$ such that the dynamical system (\ref{dina2}) takes
a simple form, which is not a trivial problem. Since we have
already analyzed the solution from the coordinate perspective, it
is a matter of simple calculation to obtain which
coordinates in $S$ provide the dynamical system
studied in section 2. Since $\bm{\partial_0}$ and $\bm{\partial_1}$
commute we can introduce coordinates $t$ and $x$ such that
$\bm{\partial_0} = 3 t/2 \partial_t$ and $\bm{\partial_1} = 3/2 \sqrt{L(x)}
\partial_x$, where $L(x)$ is an arbitrary non-vanishing function.
Parametrizing $\tilde{\Sigma}_{23}$, 
$\tilde{N}_{23}$ and $\tilde{N}_{22}$ as
\begin{eqnarray*}
\tilde{\Sigma}_{23} = \frac{\cos \alpha \cos \left (\frac{H}{2} \right)}{4},
\hspace{1cm}
\tilde{N}_{23} =  \frac{ - 2 \sin \alpha \sin Q -   \cos^2 \alpha \cos Q
\sin H}{  8  \sqrt{
\sin^2{\alpha} + \cos^2 \alpha \cos^2 \left ( \frac{H}{2}  \right )} },\\
\tilde{N}_{22} = - \frac{\cos \alpha}{4 \sin \alpha} \left [ 
4 \cos \left( \frac{H}{2} \right ) \tilde{N}_{23} +
\cos Q \sin \left(\frac{H}{2} \right )\sqrt{\sin^2 \alpha + \cos^2 
\alpha \cos^2
\left ( \frac{H}{2}  \right )} \right ],
\end{eqnarray*}
in terms of $H$ an $Q$ (which are global coordinates
on the surface $S$) the dynamical system
(\ref{dina2}) takes exactly the form (\ref{dynode}) whenever we also
choose the free function $L(x)$ as in (\ref{ele}).
Clearly, finding this
parametrization of $S$ by simply inspecting the dynamical system (\ref{dina2})
is very difficult. If we had started from the tetrad approach we would
probably have found a different parametrization of $S$ but the
description of the problem would have been essentially equivalent.
Actually, it is remarkable that both methods
lead quite naturally to a two-dimensional
dynamical system. This implies that both
the level of final understanding and the amount of the
required work are similar for the two approaches. Performing the
comparison of them is, however, very interesting because the
dimensionless variables of the tetrad formulation have a very clear physical
interpretation. Furthermore, the tetrad approach is coordinate
independent and hence it characterizes the solutions intrinsically. 
Finally, a combined use of
both methods can provide us with a powerful tool for solving more
difficult problems. In particular, the analysis of Bianchi type VII
tangent self-similar dust cosmologies has proven to be very
difficult following the coordinate approach. It is possible that using
and combining the information from the coordinate and the 
tetrad methods can help solving this case.
This matter is now under current investigation and the results will be
reported elsewhere.

\section*{Acknowledgments}

The authors wish to thank A. Koutras for pointing out that the
metric (\ref{Met2}) is self-similar and for helpful discussions.
We also thank the referee for interesting criticisms which helped improving
this work. M.M. wishes to thank
the Ministerio de Educaci\'on y Ciencia
for financial support under grant EX96 40985713.
G.H. also wishes to thank the School of Mathematical
Sciences, Queen Mary and Westfield College for the
hospitality during his stay where a great part of the
work for this paper was done.

\pagestyle{empty}
\newpage
\unitlength=1cm
\begin{figure}
\begin{picture}(10,16)
\put(-0.75,-2){
\psfig{figure=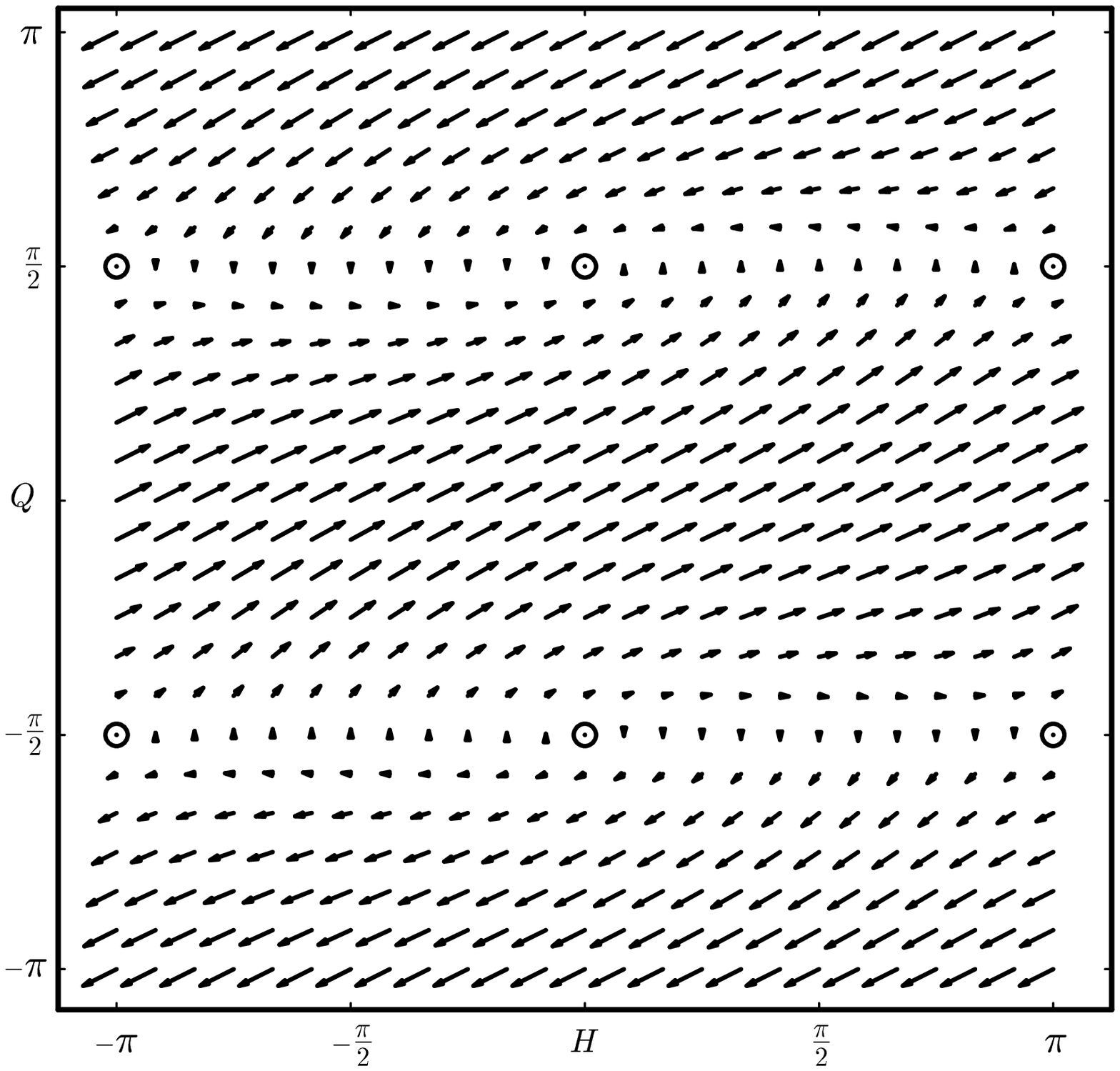,height=24cm,angle=0}}
\end{picture}
\caption{
Phase space portrait of the dynamical system (\ref{dynode}) for
$\alpha = \frac{1}{4}\pi$. The fixed points are marked by
circles
}
\end{figure}
\pagestyle{empty}
\newpage
\unitlength=1cm
\begin{figure}
\begin{picture}(10,16)
\put(-0.75,-2){
\psfig{figure=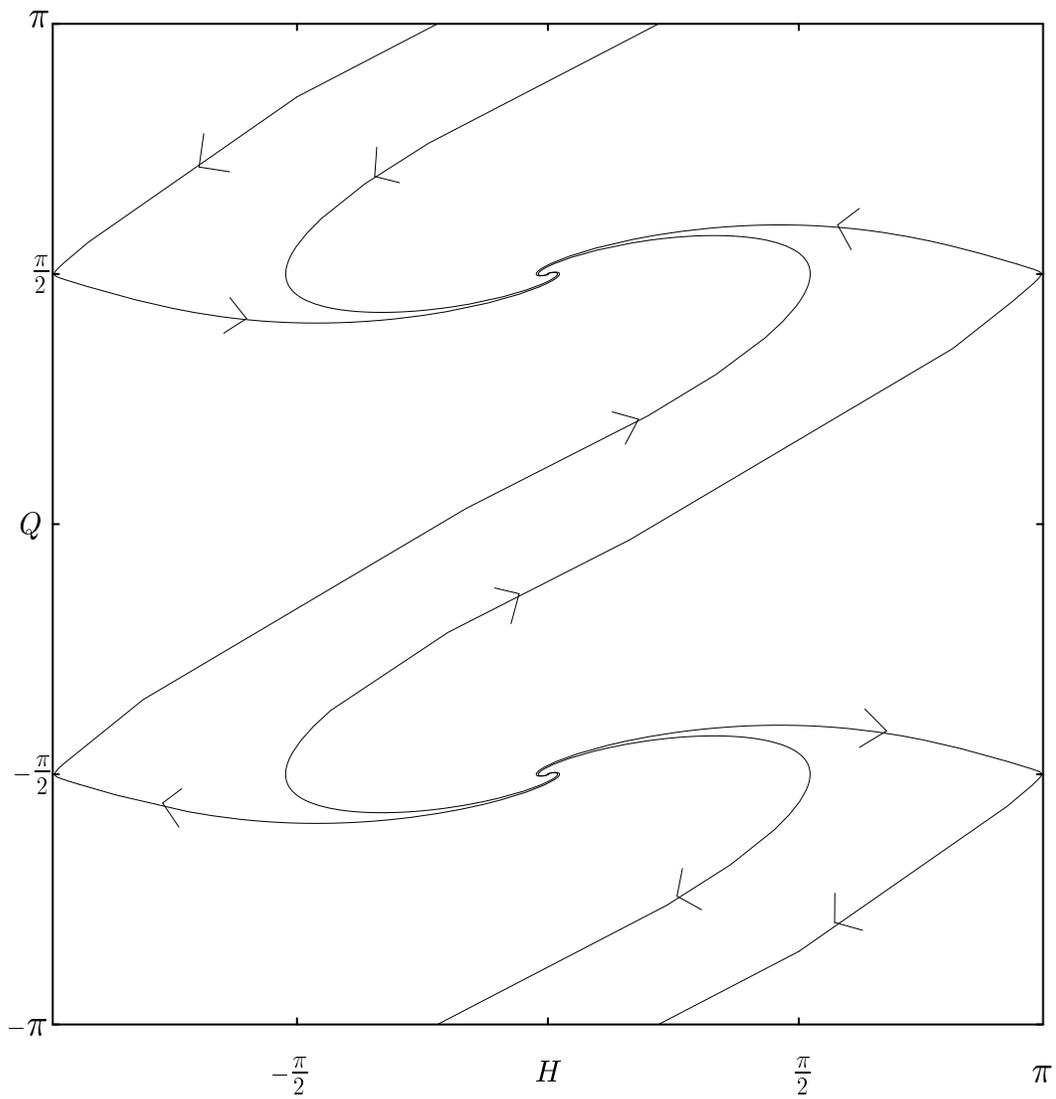,height=24cm,angle=0}}
\end{picture}
\caption{
Phase space diagram for the dynamical system (\ref{dynode}) showing
the set of trajectories starting or finishing on the saddle points
$\left ( \pi, \frac{\pi}{2} \right )$ and $\left (\pi, \frac{-\pi}{2}
\right)$ (which correspond to four points in the diagram due to the
identification of the two vertical lines). The direction
of increasing $x$ for each solution is shown. 
}
\end{figure}
\pagestyle{empty}
\newpage
\unitlength=1cm
\begin{figure}
\begin{picture}(10,16)
\put(0.75,-1){
\psfig{figure=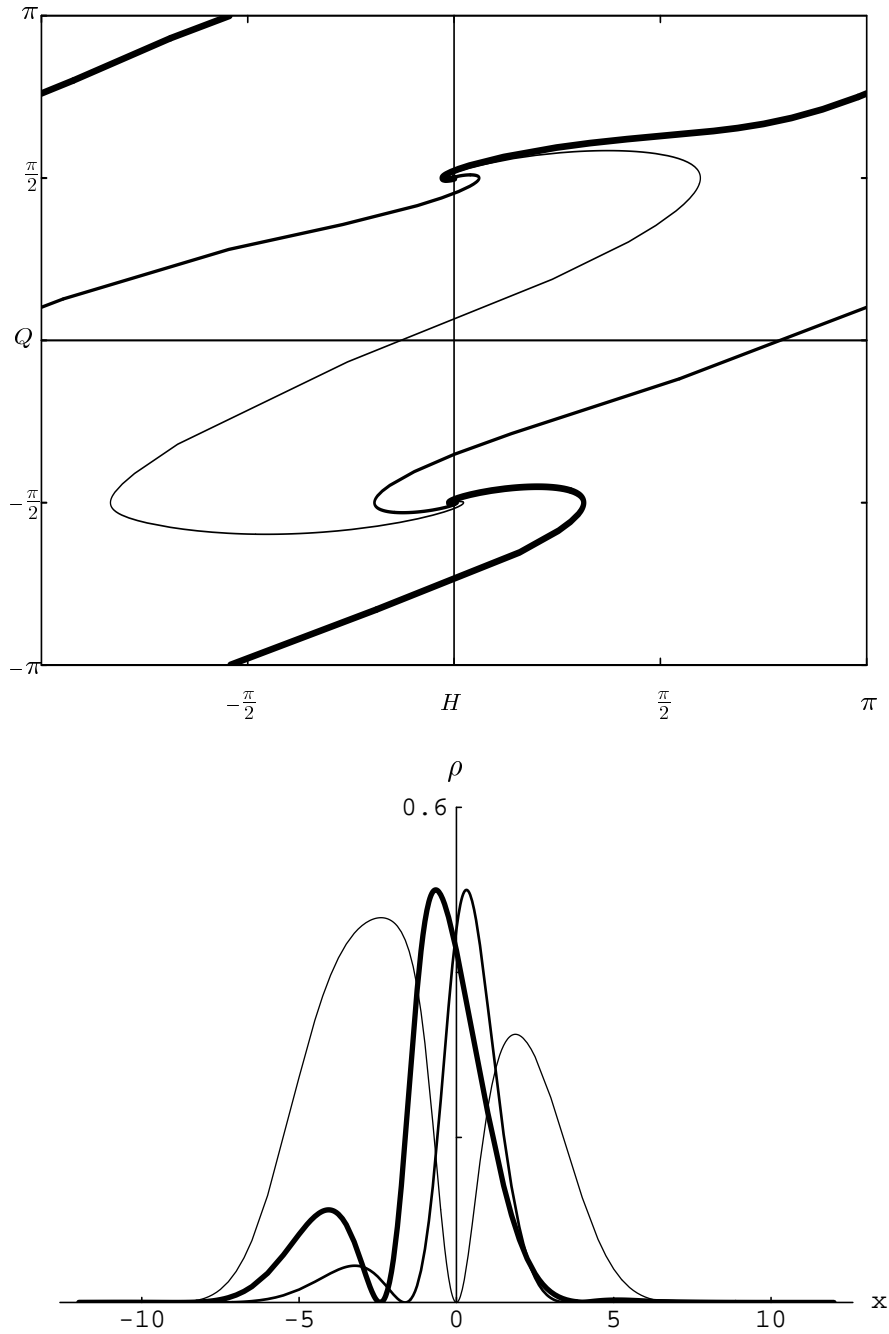,height=20cm,angle=0}}
\end{picture}
\caption{
Phase space portraits and energy-densities for three different solutions of
the dynamical system with $\alpha= \pi/4$. The thin-line solution
corresponds to the  initial values $Q(0) = \frac{1}{5} $ and $H(0) = 0$.
It never crosses the boundary of ${\cal U}$ and it does not wrap the
torus. The medium-line
solution corresponds to
the initial values $Q(0) = 0$ and $H(0) = \frac{4}{5} \pi$. This
solution wraps the torus once and the boundary
of $\cal U$ is  crossed once in the $H$ direction.
The thick-line solution corresponds to
$Q(0) = \frac{2}{3} \pi $ and $H(0) = \frac{3}{4} \pi$.
It wraps the torus once and the boundary
of $\cal U$ is  crossed twice, once in the $H$ and once in the $Q$ direction.
}
\end{figure}
\end{document}